# Towards Breath Based Diagnostics via Water-mediated Capture of Synthetic Breath Biomarkers in SERS-active Plasmonic Nanogaps

*Aditya Garg[1*], Marissa Morales[2], Aashini Shah[2], Daniel Kim[2], Ming Lei[3], Sahil Patel[4-6], Jia Dong[1], Seleem Badawy[2], Sangeeta Bhatia[5-10], Loza F. Tadesse[1,11,12*]*

[1]Department of Mechanical Engineering, Massachusetts Institute of Technology, Cambridge, MA 02139, USA

[2]Department of Health Sciences and Technology, Massachusetts Institute of Technology, Cambridge, MA 02139, USA

[3]Department of Materials Science and Engineering, Massachusetts Institute of Technology, Cambridge, Massachusetts 02139, United States

[4]Department of Pulmonary and Critical Care, Department of Medicine, Massachusetts General Hospital, Boston, Massachusetts 02124, United States

[5]Koch Institute for Integrative Cancer Research, Massachusetts Institute of Technology, Cambridge, Massachusetts 02139, United States

[6]Broad Institute of Massachusetts Institute of Technology and Harvard, Cambridge, Massachusetts 02139, United States

[7]Institute for Medical Engineering and Science, Massachusetts Institute of Technology, Cambridge, Massachusetts 02139, United States

[8]Howard Hughes Medical Institute, Cambridge, Massachusetts 02139, United States

[9]Department of Electrical Engineering and Computer Science, Massachusetts Institute of Technology, Cambridge, Massachusetts 02139, United States

[10]Department of Medicine, Brigham and Women's Hospital and Harvard Medical School, Boston, Massachusetts 02115, United States

[11]Ragon Institute of MGH, MIT and Harvard, Cambridge, MA, USA

[12]Jameel Clinic for AI & Healthcare, Massachusetts Institute of Technology, Cambridge, MA, USA

\* **Corresponding authors (email):**
Aditya Garg (aditya14@mit.edu); Loza F. Tadesse (lozat@mit.edu)

**Abstract:** Volatile organic compounds (VOCs) are valuable health indicators, with synthetic breath biomarkers offering rapid and disease-specific diagnostics. However, their <100 ppb level exhalation requires mass spectrometry, limiting clinical integration. Surface-enhanced Raman spectroscopy (SERS) offers a portable, cost-effective alternative. Yet, detecting synthetic breath biomarkers, with inherently low Raman cross-sections, at <100 ppb remains challenging. We demonstrate SERS detection down to clinically relevant 10 ppb via water-mediated trapping in hydroxylated nanoporous silica–coated plasmonic nanogaps, using pentafluoropropylamine (PFP) as a representative synthetic breath biomarker. Uniform nanogaps, with >$10^3$ electric field enhancement, were generated between a gold film and gold–silica core–shell nanoparticle assemblies using electric field-driven evaporation. Oxygen plasma treatment hydroxylated the silica, enabling water-mediated hydrogen bonding that strengthened PFP adsorption, confirmed by density functional theory. This mechanism improved SERS sensitivity by $10^4$-fold, enabling ppb level PFP detection in mouse bronchial fluid and establishing a VOC-capturing SERS platform for breath-based diagnostics.

Volatile organic compounds (VOCs) are emitted from diverse natural and anthropogenic sources, serving as valuable indicators of industrial safety, environmental quality and human health.[1,2] In particular, exhaled breath contains hundreds of trace VOCs generated by metabolic processes throughout the body, making it a promising noninvasive matrix for the rapid detection of local and systemic pathologies, including respiratory, gastrointestinal, hepatic, and renal disorders.[3,4] Among these, respiratory diseases such as lung cancer[5,6] and infectious diseases (e.g., pneumonia, and tuberculosis)[7-9] have shown particularly strong diagnostic potential due to the direct access of exhaled breath to the lungs, where many relevant metabolic processes occur. However, identifying disease-specific naturally occurring VOC biomarkers in breath remains challenging because of their extremely low concentrations and high variability influenced by individual physiology and environmental factors.[10] As a result, current clinical diagnostics still rely on invasive, resource- and time-intensive methods such as biopsies[11] and cultures of respiratory specimens[12], or indirect assessments like chest X-rays and spirometry,[13,14] which lack disease specificity, sensitivity and rely heavily on human interpretation. This underscores the clinical need for disease-specific breath-based diagnostics. To address the limitations of endogenous breath biomarkers, volatile activity-based nanosensors (vABNs) have been developed for intrapulmonary delivery, where they undergo disease-specific molecular processes to produce bio-orthogonal VOCs in exhaled breath, known as synthetic breath biomarkers.[15,16] These bio-orthogonal synthetic breath biomarkers (e.g., hydrofluoroamines, deuterated alcohols) are chemically distinct from endogenous VOCs due to the incorporation of non-natural elements like fluorine and deuterium,[15,16] enabling highly specific detection. Synthetic breath biomarkers offer high specificity and release within 10 mins of administration promising ultrasensitive and rapid diagnostics. However, they are exhaled at low (<100) parts-per-billion (ppb) levels and currently require detection via gas chromatography–mass spectrometry (GC–MS) to reach diagnostic sensitivity.[15,16] GC-MS is costly, time-intensive, and requires bulky instrumentation and trained personnel, limiting its suitability for clinical workflow integration. Thus, there is a need for a rapid, cost-effective, and ultrasensitive detection method to enable clinical translation of vABNs for breath-based diagnostics.

Surface-enhanced Raman spectroscopy (SERS) has emerged as a powerful molecular sensing approach combining the chemical specificity of Raman spectroscopy with the high sensitivity of plasmonic nanostructures.[17,18] SERS can enable rapid detection within seconds to minutes using user-friendly, portable instrumentation suitable for clinical deployment.[19] Integrating gas-trapping porous materials (e.g., metal–organic frameworks (MOFs)) into plasmonic architectures has enabled detection of volatile compounds at ppb levels.[20-22] However, such ppb level detection has been demonstrated mainly for molecules with π-electron systems (e.g., aromatic hydrocarbons) with high Raman cross sections,[20-22] limiting applicability to clinically relevant synthetic breath biomarkers such as hydrofluoroamines and deuterated alcohols, which exhibit low Raman cross sections. Thus, achieving ppb level SERS detection of synthetic breath biomarkers requires simultaneously addressing several limitations. First, the gas-trapping nanomaterials must reside within engineered plasmonic nanogaps: sub 10 nm gaps between plasmonic metals that confine intense electromagnetic fields for large SERS enhancement.[23] Second, the interaction between target VOCs and the porous nanomaterial must be energetically favorable under physiologically relevant conditions, without introducing additional linkers that can cause spectral interference or occupy the tiny nanogaps.[24] Third, the nanomaterial must provide minimal Raman background signal, unlike crystalline MOFs whose strong Raman signatures can obscure weak VOC peaks at low concentrations.[25] Finally, the plasmonic nanogap array must be uniformly assembled over large areas to enable automatic data acquisition from hundreds of nanogaps and statistically robust datasets, particularly at low VOC coverage when only a small fraction of nanogaps are occupied. Therefore, there is a need for a SERS-active platform that can address these limitations.

We have developed PlasmoSniff, consisting of gold films with assemblies of gold–silica core–shell nanoparticles formed via electric field–driven evaporation, resulting in spatially uniform resonant plasmonic nanogaps with hydroxylated Raman background free nanoporous silica for VOC trapping. The silica was hydroxylated through oxygen plasma treatment, enabling water molecules to bridge interactions between silica and engineered breath biomarkers via hydrogen bonding, substantially enhancing adsorption energy validated through density functional theory (DFT) calculations. This approach enabled detection of pentafluoropropylamine (PFP), a representative synthetic breath biomarker,[15, 16] at clinically relevant <100 ppb concentrations via SERS. Our platform enables rapid, cost-effective detection of synthetic breath biomarkers at clinically relevant concentrations, paving the way for clinically translatable, breath-based disease diagnostics.

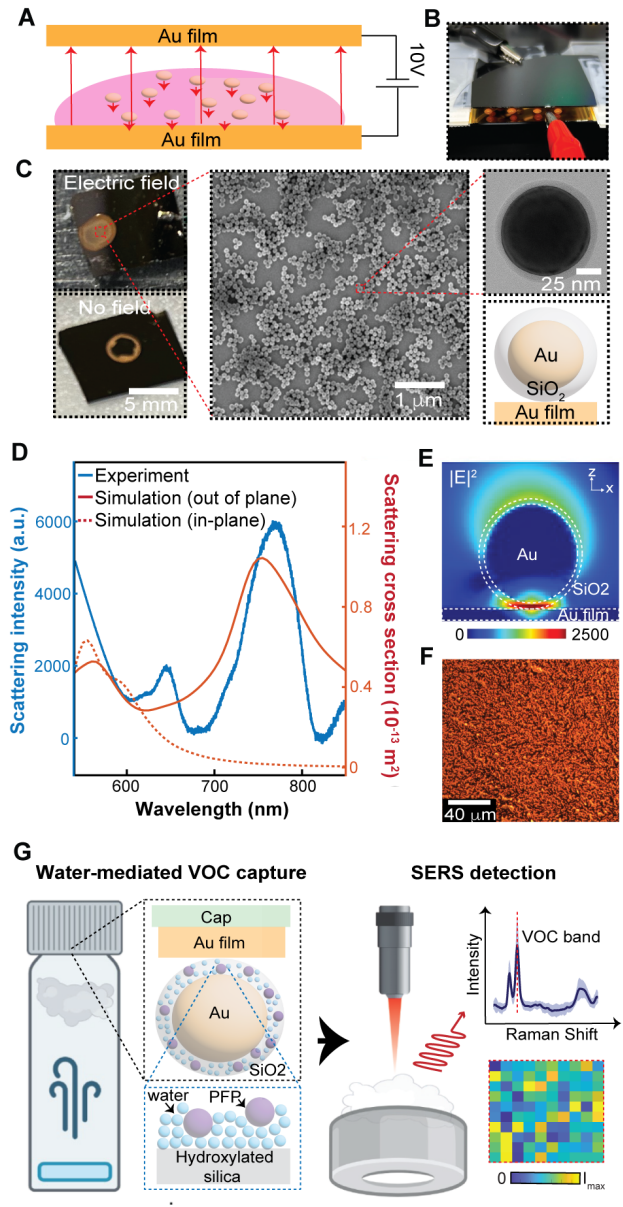

**Figure 1. Design and fabrication of PlasmoSniff for SERS-based synthetic breath biomarker detection:** (A) Schematic illustration and (B) experimental setup of the electric field–driven nanoparticle drying method for uniform nanoparticle deposition. (C) Images of plasmonic nanogap chips at multiple scales. (Left-top) Photograph showing uniform nanoparticle coverage on the gold surface achieved through electric field-driven drying, in contrast to (Left-bottom) the ring-like pattern observed with conventional drying. (Middle) SEM images display the nanoparticle distribution on the gold surface, while a (Right-top) zoomed-in TEM image reveals the nanoporous silica coating on the gold nanoparticles. (Right-bottom) A schematic representation of the nanoparticle-on-gold film architecture is also shown. (D) Optical properties of the PlasmoSniff chip including the experimental dark field scattering spectrum and FDTD-simulated scattering spectra under in plane and out of plane polarized excitation. (E) FDTD-simulated electric field intensity ($|E|^2$) map under 785 nm excitation showing $|E|^2>10^3$. (F) Dark field scattering image showing uniform spatial distribution of nanogap scattering. (G) Schematic illustration of the workflow for water-mediated VOC capture and SERS detection using PlasmoSniff.

We designed and fabricated the PlasmoSniff platform to obtain spatially uniform VOC-trapping plasmonic nanogaps resonant with the 785 nm SERS excitation wavelength. To achieve uniform nanoparticle deposition on gold films, conventional approaches often rely on either modifying the solvent composition (e.g., using low surface tension organic solvents)[22] or altering the nanoparticle surface capping (e.g., via the Langmuir-Blodgett technique).[26] However, both methods introduce organic species that can produce interfering Raman background signals, masking the characteristic peaks of the target analytes, particularly at low ppb-level concentrations. Here, we employ an organics-free, electric field-driven drying method to uniformly deposit negatively charged, silica-coated gold nanoparticles (ζ-potential ≈ –25 mV) suspended in water onto a positively charged gold substrate (Figures 1A-B). A potential difference of 10 V is applied between two parallel gold plates separated by 0.5 cm, generating a uniform downward electric field of approximately $2 \times 10^3$ V·m$^{-1}$. The resulting electrostatic force drives the nanoparticles downward toward the substrate, counteracting the outward capillary-driven convective flow, thereby suppressing the coffee ring effect.[27]

As shown in Figure 1C, electric field-driven drying yields a uniform nanoparticle film across the gold surface, in contrast to the ring-like deposition pattern observed with conventional evaporation. Scanning electron microscopy (SEM) reveals dense and homogeneous nanoparticle coverage with micron-scale aggregates. The limited aggregation that does occur likely arises from capillary attractions between adjacent particles during the final stages of evaporation,[28] which are insufficiently counterbalanced by electrostatic repulsion, particularly in the absence of organic capping agents. Zoomed-in transmission electron microscopy (TEM) images reveal a ~5 nm thick, dense, amorphous silica layer on the 100 nm gold nanoparticles, with sub nm pores ideal for gas trapping. These nanoparticles are assembled on the gold film to form a particle-on-mirror architecture, enabling intense field confinement within the nanogaps for SERS enhancement,[29] with the nanogap distance governing the plasmonic resonance controlled by the thickness of the silica layer.[30]

To investigate the plasmonic behavior of this architecture, finite-difference time-domain (FDTD) simulations were performed under two excitation schemes where the incident electric field was either perpendicular (out of plane) or parallel (in plane) to the gold film (Figure 1D). Under out of plane excitation, the simulated scattering spectrum exhibited a dominant resonance band in the 740-800 nm region, corresponding to a vertically coupled plasmonic mode that aligns well with the 785 nm SERS excitation wavelength.[31, 32] This out of plane mode generated intense electromagnetic field enhancement ($|E|^2 > 10^3$) within the silica layer, optimal for SERS enhancement (Figure 1E). As the silica spacer thickness increased, this out of plane mode exhibited a gradual blueshift, leading to reduced spectral overlap with the 785 nm excitation (Figure S1). In contrast, in plane excitation produced weak scattering due to destructive interference between the horizontally-oriented nanoparticle dipole and its image dipole in the gold film.[31, 32] Experimentally, the dark field scattering spectrum showed a strong band centered at ~770 nm, in close agreement with the simulated out of plane response. Although the experimental illumination was nominally at normal incidence, the conical nature of the focused beam through the objective introduces an out of plane polarization component, enabling excitation of the vertically coupled mode. The dark field images further revealed a dense and uniform nanoparticle coverage with a consistent reddish hue, indicating homogeneous nanogap formation across the film (Figure 1F). Such large-area uniformity enables large-area SERS mapping, in contrast to conventional single-spot measurements in nanogap-based configurations, thereby facilitating automated scanning and the acquisition of statistically robust datasets.

Prior to use, the PlasmoSniff chips were treated with oxygen plasma which serves two purposes: i) to remove organic contaminants through bombardment with reactive oxygen species;[33] and ii) to increase the hydrophilicity of the silica by replacing the surface hydrocarbons with oxygen-containing functional groups (e.g., hydroxyl).[33, 34] After plasma treatment, the substrate's SERS spectrum exhibited minimal interfering Raman bands, consistent with the removal of organic contaminants and the inherently low Raman background from amorphous silica (Figure S2).[35] Figure 1G illustrates the workflow for SERS-based detection of VOCs using PlasmoSniff. The chips were taped to the underside of a sealed tube cap and exposed to volatilized PFP for 2 hours at 70 °C under controlled humidity. During incubation, PFP adsorbs onto the hydroxylated silica nanogaps through water-mediated hydrogen bonding. Subsequently, SERS area scans were performed over 200x200 µm² regions enabling mapping of characteristic Raman peaks and PFP quantification.

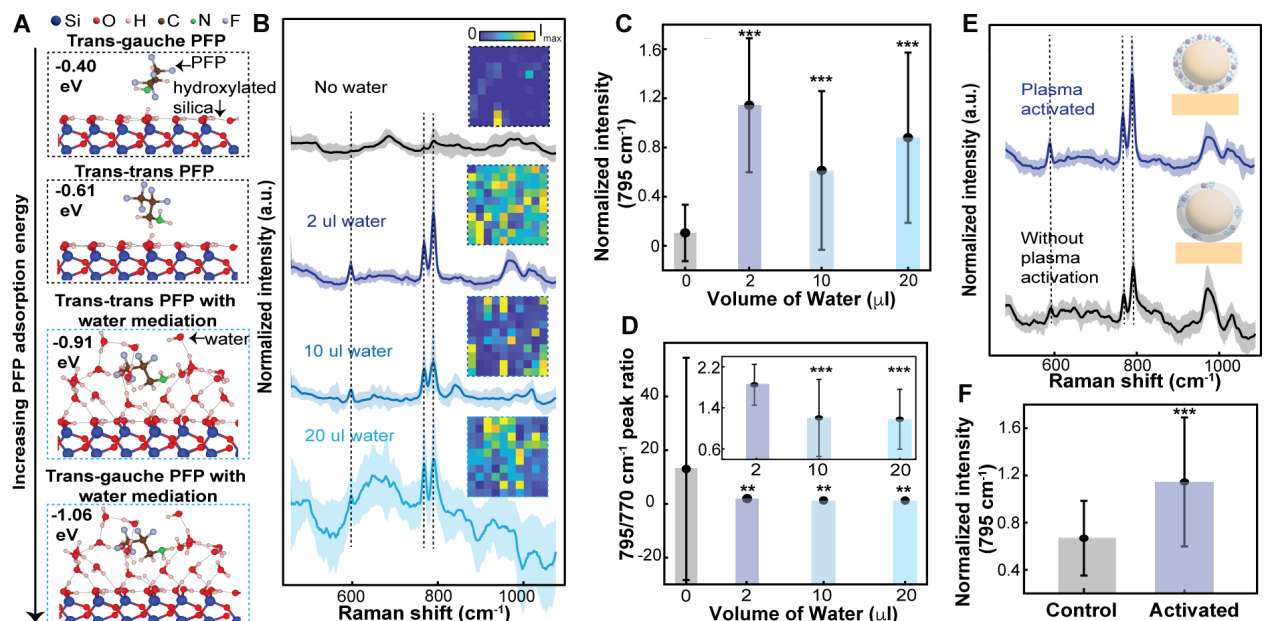

**Figure 2: Water-mediated PFP adsorption and SERS detection using PlasmoSniff.** (A) DFT-calculated molecular structures and corresponding adsorption energies of the trans–trans and trans–gauche conformers of PFP on hydroxylated silica showing improved adsorption in the presence of water and enhanced stabilization of the trans-gauche conformer. (B) SERS spectra of PFP volatilized in the presence of varying water volumes (0, 2, 10, and 20 µL). Shaded regions indicate standard deviations, and black lines mark the characteristic PFP peaks at 595, 770, and 795 cm$^{-1}$. The inset shows corresponding SERS maps of the 795 cm$^{-1}$ PFP peak. (C) Quantitative comparison of the 795 cm$^{-1}$ PFP peak as a function of water volume showing increased signal intensity in the presence of water. (***$p < 0.001$ vs 0 µL water case (two-sample t-test); n=100; error bars represent standard deviations). (D) Ratio of the 795 cm$^{-1}$ and 770 cm$^{-1}$ PFP peaks as a function of water volume showing decrease in peak ratio with water level. (***$p < 0.001$ and **$p < 0.01$ vs 0 µL water case (two-sample t-test); n=100; error bars represent standard deviations). (E) SERS spectra of PFP in the presence of 2 µL water obtained using oxygen plasma-activated and non-activated samples, and (F) corresponding quantitative comparison of the 795 cm$^{-1}$ peak intensities showing improved signal intensity upon plasma activation. (***$p < 0.001$ vs control (two-sample t-test); n=100; error bars represent standard deviations).

We hypothesized that interfacial water molecules could enhance the adsorption of PFP onto hydroxylated silica surfaces. Under humid conditions, water molecules form strong hydrogen bonds with surface silanol groups, giving rise to a dense interfacial water layer within the silica network.[36, 37] These water molecules can also hydrogen bond with the amine moiety of PFP, thereby stabilizing PFP-water-silica complexes within the nanogaps. To test this hypothesis, we performed DFT calculations. In the gas phase, PFP exhibits two major stable conformers: *trans–trans* and *trans–gauche*.[38] On the hydroxylated $SiO_2$ (001) surface in the absence of water, the adsorption energies of the *trans–trans* and *trans–gauche* conformers are −0.61 and −0.40 eV, respectively, indicating that the *trans–trans* structure is energetically favored (Figure 2A). Upon introducing a water layer, the adsorption energies increase in magnitude to −0.91 and −1.06 eV for the *trans–trans* and *trans–gauche* conformers, respectively (Figure 2A). This demonstrates that the presence of water substantially strengthens PFP adsorption by forming hydrogen bonds that bridge the PFP molecule and the $SiO_2$ surface. Notably, the *trans–gauche* conformer becomes more stable than the *trans–trans* form in the hydrated environment, consistent with the stabilization of polar conformers in polar media through dipole-solvent interactions.[39]

To experimentally probe this water-mediated adsorption mechanism, varying volumes of water (2, 10, and 20 µL) were introduced into sealed tubes containing PFP ($10^7$ ppb) to generate controlled humidity levels. SERS spectra showed pronounced enhancement of PFP peaks at 595 $cm^{-1}$ ($CF_3$ antisymmetric deformation), 770 $cm^{-1}$ ($CF_3$ symmetric stretching modes in *trans-gauche* conformer), and 795 $cm^{-1}$ $cm^{-1}$ ($CF_3$ symmetric stretching modes in *trans-trans* conformer) in the presence of water compared to the dry control (Figure 2B).[38] Quantitative analysis revealed an approximately 20-fold increase in the intensity of the 795 $cm^{-1}$ peak in the presence of 2 µL of water relative to the dry condition (Figure 2C). The signal intensity of $10^7$ ppb PFP measured under dry conditions approximately matches that obtained from $10^3$ ppb PFP in the presence of 2 µL of water (Figure S3), corresponding to a four order of magnitude enhancement in detection sensitivity. Interestingly, a modest 1.3-1.8-fold decrease in signal is observed as the water volume increased to 10 µL and 20 µL (Figure 2C). The standard deviation of the 795 $cm^{-1}$ peak intensity also increased from 0.54 to 0.64 and 0.69 as the water volume increased, indicating greater heterogeneity in the SERS response at higher humidity levels. We attribute this behavior to the accumulation of excess water molecules unassociated with PFP within the silica nanogaps, diminishing signal intensity and increasing spatial variability. Spatial SERS maps of the 795 $cm^{-1}$ peak further corroborated these findings, showing a pronounced increase in high-intensity pixels in the presence of water, yet progressively reduced signal uniformity at higher water volumes (Figure 2B inset).

Interestingly, a marked decrease in the intensity ratio between the 795 $cm^{-1}$ (*trans–trans*) and 770 $cm^{-1}$ (*trans–gauche*) peaks was observed upon the addition of water (Figure 2D). This experimental trend is consistent with our DFT calculations, which revealed that water mediation increases the adsorption energy of the *trans–gauche* conformer (from –0.40 eV to –1.06 eV) more significantly than that of the *trans–trans* conformer (from –0.61 eV to –0.91 eV). (Figure 2A). This result further supports the proposed PFP-water interaction mediated adsorption mechanism. Finally, we evaluated the effect of oxygen plasma induced surface hydrophilization on water-mediated PFP adsorption by comparing freshly plasma treated and aged controls that lost hydrophilicity after 3 days of ambient exposure.[40] The results showed that oxygen plasma activated samples exhibited approximately a 2-fold increase in SERS signal intensity, underscoring the critical role of surface hydroxylation and enhanced hydrophilicity in facilitating water-mediated PFP adsorption within the nanogaps (Figures 2E-F).

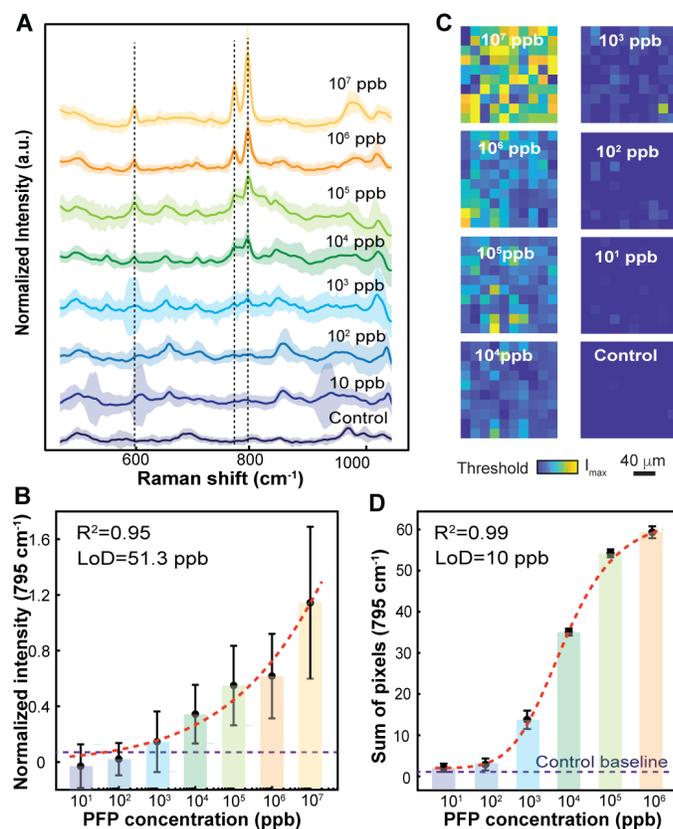

**Figure 3: SERS detection of volatilized PFP using PlasmoSniff.** (A) SERS spectra of volatilized PFP across concentrations ranging from $10^7$ to $10^1$ ppb showing PFP Raman bands. Shaded regions indicate standard deviations, and black dotted lines mark the characteristic PFP peaks at 595, 770, and 795 cm$^{-1}$. (B) Quantitative analysis of the 795 cm$^{-1}$ PFP peak intensity across the concentration range, showing a mean-based detection limit of 51.3 ppb. (n=100; Error bars represent standard deviations; the red curve shows the sigmoidal calibration fit; the purple dashed line shows the baseline for LoD calculation defined as mean plus three times the standard deviation of the control). (C) Spatial SERS intensity maps of the 795 cm$^{-1}$ PFP peak, showing positive pixels down to 10 ppb. Threshold is defined as mean plus two times the standard deviation of the control. (D) Cumulative pixel intensities from digital SERS analysis, showing a detection limit of 10 ppb. (n=2; Error bars represent standard deviations; the red curve shows the sigmoidal calibration fit; the purple control baseline reflects the cumulative pixel intensities from the control sample).

After optimizing the device design and adsorption chemistry, we evaluated the detection sensitivity of PFP. Figures 3A and S4 show the SERS spectra of volatilized PFP at concentrations ranging from $10^7$ ppb to 10 ppb in the presence of 2 µL of water, revealing distinct PFP bands at 595, 770 and 795 cm$^{-1}$. For quantification, the 795 cm$^{-1}$ peak intensity was fitted to a sigmoidal calibration curve, yielding an R² value of 0.95 (Figures 3B). Based on this calibration for the 795 cm$^{-1}$ peak, the limit of detection (LoD) for volatile PFP was determined to be 51.3 ppb. Interestingly, SERS mapping of the 795 cm$^{-1}$ band showed numerous positive pixels compared to the control sample even at 10 ppb (Figure 3C). At such low analyte concentrations, only a limited subset of plasmonic nanogaps captures detectable PFP molecules, resulting in spatially heterogeneous SERS intensities near the noise floor. This variability makes conventional mean-based quantification unreliable. To overcome this limitation, we implemented a digital SERS analysis, which evaluates individual pixel intensities by excluding those below a defined threshold and summing the remaining ones.[41, 42] The threshold was set as the mean plus two standard deviations of the negative control, ensuring that 90% of control pixels fall below this value and minimizing background contributions. The resulting summed pixel intensities followed a sigmoidal calibration curve ($R^2 = 0.95$) and remained above the control baseline down to 10 ppb (Figure 3D). Therefore, digital SERS analysis improved the LoD from 51.3 ppb to 10 ppb. Notably, this sensitivity surpasses the concentrations of exhaled synthetic breath biomarkers detected in infected mice (~15-30 ppb) following the administration of inhalable vABNs at 100 μM levels,[15, 16] underscoring the clinical relevance of the developed platform.

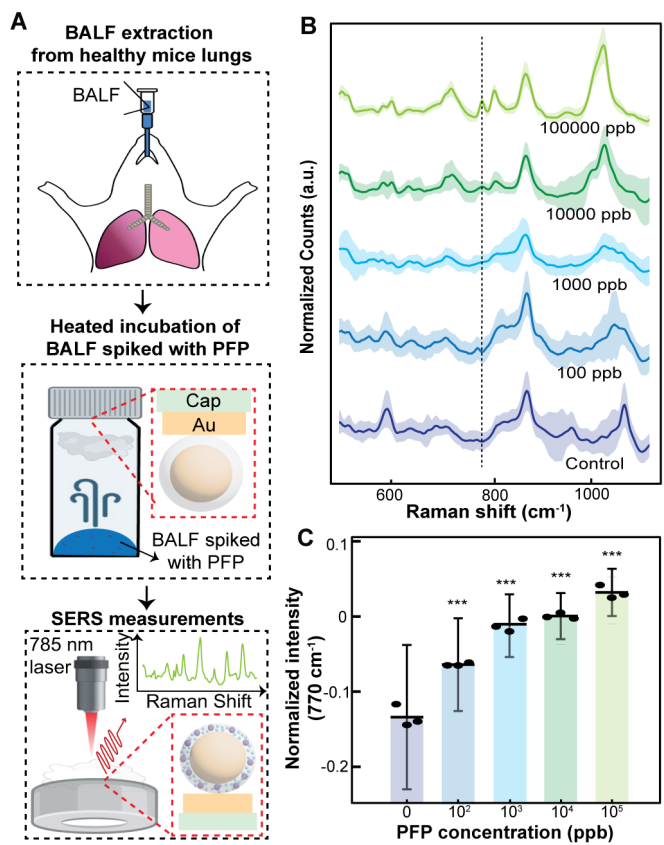

**Figure 4: SERS detection of PFP in mice bronchoalveolar lavage fluids (BALF) using PlasmoSniff.** (A) Schematic of the experimental workflow for SERS detection of PFP spiked into mice BALF using PlasmoSniff. (B) SERS spectra of volatilized PFP in BALF across concentrations from $10^2$ ppb to $10^5$ ppb. Shaded regions indicate standard deviations, and the black dotted line marks the characteristic PFP peak at 770 cm$^{-1}$, located in an interference-free region. (C) Quantitative analysis of the 770 cm$^{-1}$ PFP peak intensity showing statistically significant signals down to 100 ppb. (bars show overall mean ± standard deviation; n=300, dots represent individual mouse means, 3 mice per group; ***$p < 0.001$ vs. control (two-sample t-test)).

Next, we investigated whether our detection scheme could be implemented in simulated breath conditions using spiked BALF samples, which are recovered from the mice lungs. BALF comprises various VOCs which would be released in breath, including alcohols that strongly interact with water and aromatic compounds (e.g., toluene) with high Raman cross sections.[43-45] These endogenous VOCs can generate interfering SERS signals on our platform, potentially masking the target biomarker signals. To demonstrate SERS detection in the presence of the breath background, BALF was extracted from healthy mouse lungs, spiked with PFP, and incubated at 70 °C in a sealed tube containing the plasmonic chips, followed by SERS area scanning (Figure 4A). Figure 4B shows the SERS spectra of BALF samples spiked with PFP at concentrations ranging from 100 to 10,000 ppb. The spectra display characteristic PFP peaks at 795 and 770 $cm^{-1}$, superimposed on background features attributed to VOCs released by the BALF matrix. Because the 795 $cm^{-1}$ peak partially overlaps with a background band, the 770 $cm^{-1}$ peak was used for quantitative analysis. As evident in Figure 4C, the intensity of the 770 $cm^{-1}$ peak increased significantly compared to controls across the entire concentration range, demonstrating that this approach can reliably detect engineered volatile biomarkers at ppb levels even in the presence of complex background noise expected in breath samples.

In summary, this work demonstrates that spatially uniform plasmonic nanogaps formed between a gold film and assemblies of Au-SiO$_2$ core-shell nanoparticles effectively enable SERS detection of synthetic breath biomarker PFP at ppb levels via a water-mediated adsorption mechanism. DFT simulations reveal that water molecules mediate and stabilize interactions between PFP and oxygen plasma treated hydroxylated silica via hydrogen bonding. Subsequent, experimental results demonstrated a four order of magnitude improvement in detection sensitivity attributed to the water-mediated adsorption. Our approach achieved a clinically relevant detection limit of 10 ppb needed for respiratory disease diagnostics. To the best of our knowledge, this is the lowest reported SERS limit of detection for VOCs without π-electron systems responsible for high Raman cross sections. We further mimic *in vivo* breath conditions, successfully detecting 100 ppb PFP spiked into mouse BALF, highlighting the clinical relevance and applicability of this strategy for respiratory disease diagnostics. To further improve detection sensitivity in breath, future work will focus on flow systems capable of sequentially injecting stored breath samples onto the PlasmaSniff chips, allowing accumulation of VOCs within the nanogaps. For clinical translation, in vivo validation will be necessary to monitor exhaled volatile reporters generated by protease cleavage in infected patients, using an integrated platform for collection and incubation of exhaled breath VOCs prior to SERS analysis. The combination of PlasmaSniff's sensitivity and the selectivity of synthetic breath biomarkers could facilitate rapid, low-cost breath-based disease diagnostics compatible with clinical workflows. Beyond diagnostics, this platform could be adapted to detect a broad range of environmental and industrial VOCs capable of hydrogen bonding with water such as amines, alcohols, and ketones,[46, 47] supporting real-time hazard monitoring.

**Supplementary Information**

Methods, simulated scattering spectra with different silica thicknesses, Background SERS spectra upon oxygen plasma cleaning, comparison of PFP SERS spectra under dry and humid conditions, technical replicate for PFP quantification.

Notes

The authors declare no competing financial interest.

Author Contributions

A.G. and L.F.T. conceptualized the PlasmoSniff platform. A.G. performed the FDTD simulations, device fabrication, and optical and structural characterization. M.L. performed DFT calculation. A.G., M.M., A.S., and J.D. developed the experimental setup for VOC incubation and SERS measurements. D.K. and S.P. extracted mice BALF samples and provided clinical guidance along with S.B. A.G. conducted the experiments and data analysis, with support from M.M., A.S., and S.B. A.G. wrote the paper with contributions from L.F.T., M.M., and A.S. and feedback from all authors.

Acknowledgments

The authors gratefully acknowledge the funding from Open Philanthropy (grant number 3860485). D.K. was supported by award Number T32GM144273 from the National Institute of General Medical Sciences. The content is solely the responsibility of the authors and does not necessarily represent the official views of the National Institute of General Medical Sciences or the National Institutes of Health. The authors also thank Dr. Heather E. Fleming and Dr. Srilakshmi Premachandran for feedback on the manuscript, and Dr. Yong Zhang for assistance with TEM imaging.

Supplementary information:

**Methods:**

**Fabrication of PlasmoSniff Chips.** Au–SiO$_2$ core–shell nanoparticles (100 nm Au core, 5 nm SiO$_2$ shell) were obtained from Nanopartz Inc. (uncapped, solvent: deionized water). The nanoparticles were resuspended in fresh DI water at a concentration of 6 × 10$^{11}$ particles mL$^{−1}$ prior to use. To fabricate the electrodes, silicon wafers were coated with 5 nm Ti (adhesion layer) and 50 nm Au via electron-beam evaporation. Several 1 µL droplets of the nanoparticle suspension were deposited across the gold-coated wafer surface. In a custom-built assembly, a second Au-coated wafer was placed on top, separated by ~0.5 mm spacers (glass slides), and a 10 V potential was applied between the two electrodes, with the positive terminal connected to the nanoparticle-coated electrode. The nanoparticle solution was allowed to evaporate under ambient conditions in a fume hood for 2 h, forming uniform nanogap assemblies. The resulting chips were exposed to oxygen plasma (50 W, 50 sccm O$_2$) for 1 min to remove organic contaminants, rinsed with isopropanol, and then subjected to a second oxygen plasma treatment for 30 min to eliminate residual organics and activate the surface with hydroxyl groups. The chips were subsequently diced and affixed to the caps of 20 mL scintillation vials for further analysis. All chips were used within 1 h of plasma activation. For control experiments without plasma activation (Figures 2E–F), plasma-treated chips were left in a fume hood for 3 days, resulting in loss of hydrophilicity due to environmental aging.[1]

**FDTD simulations.** Finite-difference time-domain (FDTD) simulations were performed using commercial software (FDTD Solutions, Ansys–Lumerical Inc.) to numerically compute the far-field scattering spectra and near-field electromagnetic distributions of the PlasmoSniff chips. The optical constants for Au were taken from Johnson and Christy, and the refractive index of SiO$_2$ was set to 1.5. The Au core diameter was fixed at 100 nm, while the SiO$_2$ shell thickness was varied. A uniform mesh with 1 nm resolution in all three spatial directions was used. Perfectly matched layer (PML) boundary conditions were applied in the x, y, and z directions, and the background refractive index was set to 1.0.

**Dark field scattering measurements.** Dark-field scattering spectra were recorded using an upright confocal microscope (alpha300 RSA+, WITec) coupled to a halogen lamp (HAL 100, Zeiss) and a CCD detector (Oxford Instruments). Spectra were collected through a 50× dark-field objective (numerical aperture=0.75) using a 300 grooves mm$^{−1}$ grating. The scattering background was subtracted using spectra acquired from adjacent nanoparticle-free regions. The exposure time was set to 5s.

**Density functional theory (DFT) calculations.** All spin-polarized DFT calculations were performed using the Vienna *Ab initio* Simulation Package (VASP 5.4.4).[2] The electronic exchange–correlation interactions were described within the generalized gradient approximation (GGA) using the Perdew–Burke–Ernzerhof (PBE) functional.[3] A plane-wave energy cutoff of 500 eV was applied throughout all calculations. Electronic and ionic relaxations were considered converged when the total energy change was below 10$^{−6}$ eV and the residual atomic forces were smaller than 0.02 eV·A$^{−1}$. Long-range dispersion interactions were accounted for using Grimme's DFT-D3 correction.[4] The Brillouin zone was sampled with a 3 × 3 × 1 Monkhorst–Pack *k*-point mesh. The hydroxylated SiO$_2$ surface was modeled using an *α*-SiO$_2$ (001) slab containing four atomic layers of Si. To passivate the dangling bonds, both the upper and lower surfaces were terminated with hydrogen atoms on the unsaturated oxygen sites.[5, 6] During geometry optimization, the bottom two layers were fixed while the top two layers were fully relaxed. To simulate the effect of water vapor, nineteen $H_2O$ molecules were uniformly distributed above the hydroxylated $SiO_2$ (001) surface.

The adsorption energy of the pentafluoropropylamine (PFP) molecule on the $SiO_2$ surface was evaluated using: $\Delta E = E_{PFP@SiO2} - E_{PFP} - E_{SiO2}$, where $E_{PFP@SiO2}$, $E_{PFP}$, and $E_{SiO2}$ correspond to the total energies of the adsorbed system, the isolated PFP molecule, and the clean $SiO_2$ surface, respectively.

**Heated incubation of PFP.** Liquid PFP at the desired concentration was drop-cast into 20 mL scintillation vials along with the designated water content. The vials were sealed with caps containing the plasma-activated nanogap chips and further sealed with parafilm. The sealed vials were incubated at 70 °C for 2 h before measurements.

**BALF experiments.** Bronchoalveolar lavage fluid (BALF) was collected from mice as described previously.[7] Briefly, mice were first euthanized, and then an incision was made into the trachea, followed by needle puncture to perform lavage with a single 1 ml injection of PBS via a catheter-fitted syringe. BALF was centrifuged at 8,000xg for 5 minutes at 4 °C to remove cells and debris. PFP at defined concentrations was spiked into BALF and 2 ul was drop-cast into 20 mL scintillation vials. The vials were sealed and incubated as described above. The experiments were repeated using BALF collected from three independent mice.

**SERS measurements of PFP.** SERS measurements were performed using an upright confocal Raman microscope (alpha300 RSA+, WITec Inc.) equipped with a 785 nm diode laser. The backscattered photons were dispersed by a spectrometer (UHTS300, WITec Inc.) and detected using a CCD camera (DU401A, Oxford Instruments). Spectra were collected through a 50× objective (numerical aperture=0.55) using a 300 grooves mm$^{-1}$ grating. Acquisition parameters were as follows: 2 mW laser power, 2 s integration time, and 100 pixels per scan over a 200 μm × 200 μm area.

**Data analysis.** Following acquisition, spectra were processed using WITec Project v4.1 software. A fourth-order polynomial baseline correction, cosmic ray removal, and Savitzky–Golay smoothing (window size = 7, polynomial order = 3) were applied. For Figures 2–3, SERS intensities were normalized to the $SiO_2$ peak at 260 cm$^{-1}$.[8] In BALF experiments (Figure 4), spectra were normalized to the plasmonic electronic Raman scattering (ERS) signal at 81 cm$^{-1}$ due to interference in the 260 cm$^{-1}$ region. The ERS signal, arising from metal-based plasmon-enhanced electronic scattering, served as an internal standard for spatial and temporal calibration of molecular SERS signals from analytes within the same hotspots, enabling quantitative analysis.[9]

Statistical comparisons between groups were conducted using independent two-sample Student's t-tests in MATLAB, and p-values were used to assess significance. Calibration curves were fitted using a logistic sigmoidal model: $y = \frac{a}{1+\exp(-b*(x-c))}$, where x = $\log_{10}$(concentration of PFP in ppb) and y=normalized Raman peak intensity.

To calculate the mean-based limit of detection (LoD), Mean$_{control}$ + 3*SD$_{control}$ was used as the baseline. Mean$_{control}$ and SD$_{control}$ are the average intensity and standard deviation, respectively, of the 795 cm$^{-1}$ PFP peak in control samples without PFP. For digital SERS analysis, the SERS maps at different PFP concentrations were converted into a binary format (0 or 1) based on whether the pixel's intensity exceeded a predefined threshold. The threshold was still set as $Mean_{control} + 2*SD_{control}$. Each pixel of the digitized map was multiplied by each corresponding pixel of the original SERS map, thus generating a digital SERS map. The sum of the pixel intensities for each map at various concentrations was used for LoD quantification, with the baseline determined from the summed pixel intensity of control samples.[10]

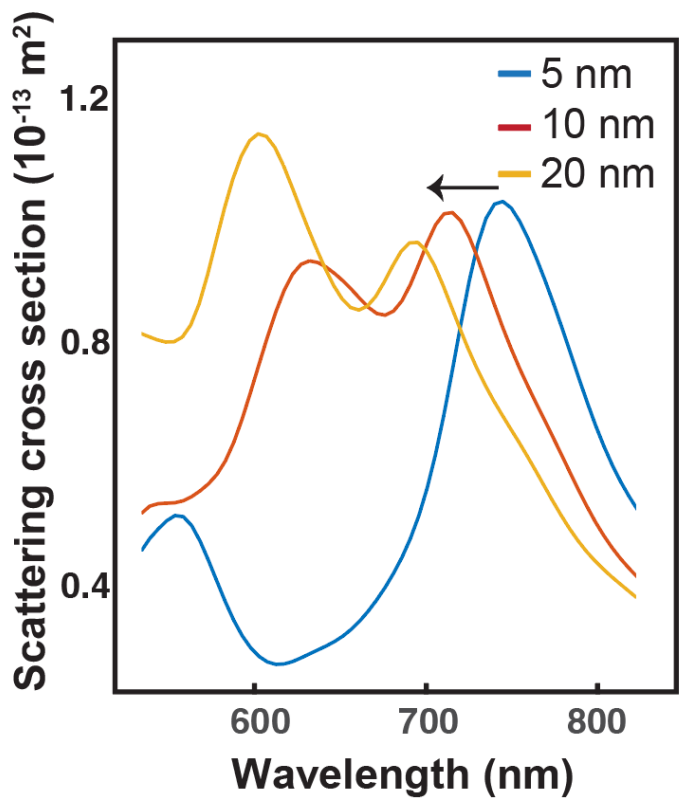

**Figure S1:** Simulated scattering spectra of the particle on mirror architecture with 5, 10 and 20 nm silica spacer thicknesses under out of plane electric field excitation.

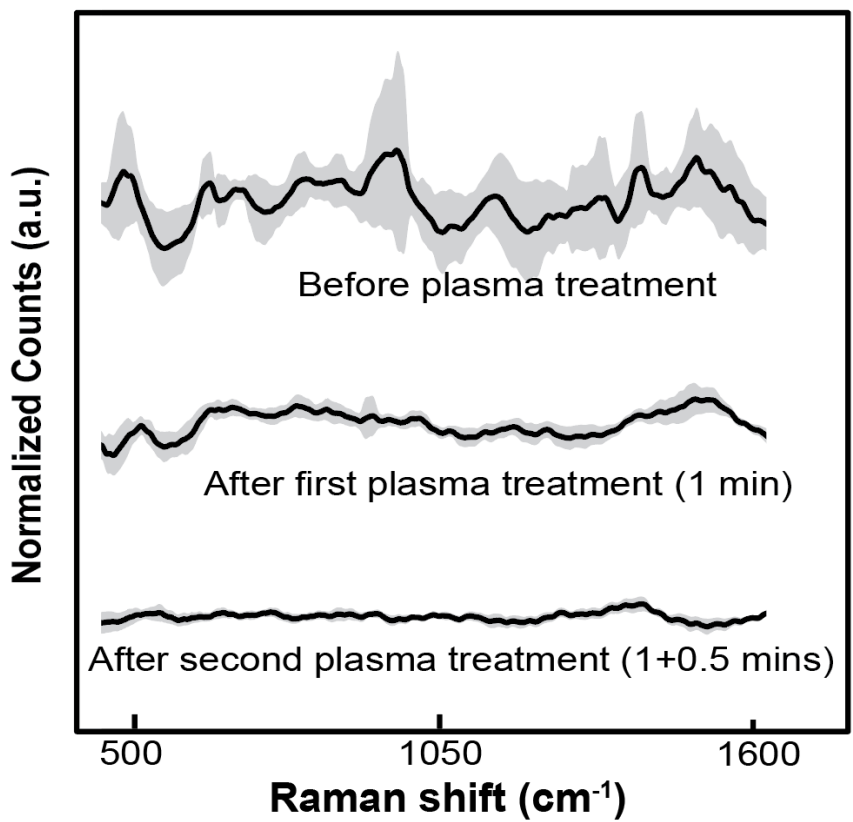

**Figure S2:** SERS spectra of the plasmonic nanogap chips before plasma treatment, after the first round of oxygen plasma treatment (1 min), and after the second round of oxygen plasma treatment (1+0.5mins).

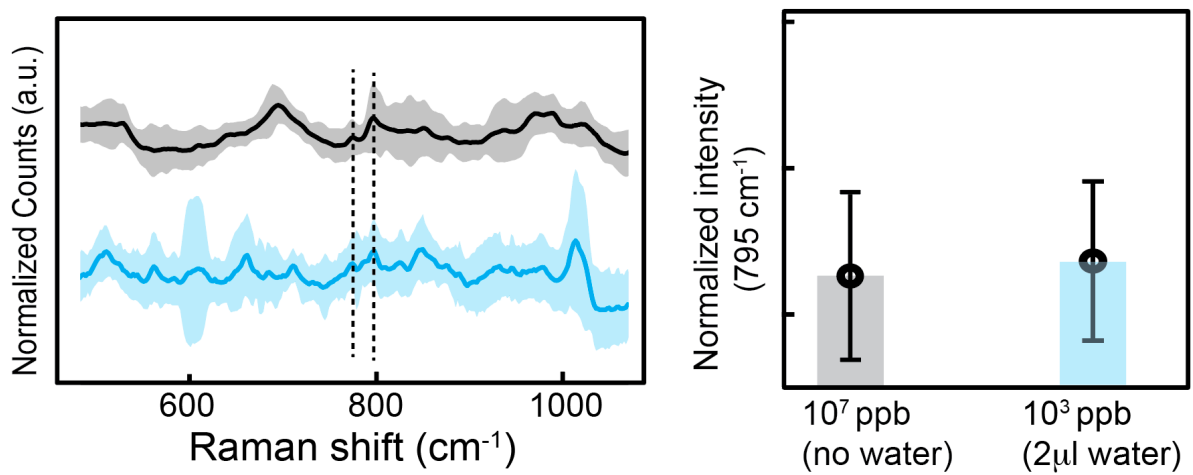

**Figure S3:** A) SERS spectra of the $10^7$ ppb PFP under dry conditions and $10^3$ ppb PFP in the presence of 2ul water. Quantitative comparison of the 795 cm$^{-1}$ peak from $10^7$ ppb PFP under dry conditions and $10^3$ ppb PFP in the presence of 2ul water. (Error bars represent standard deviations)

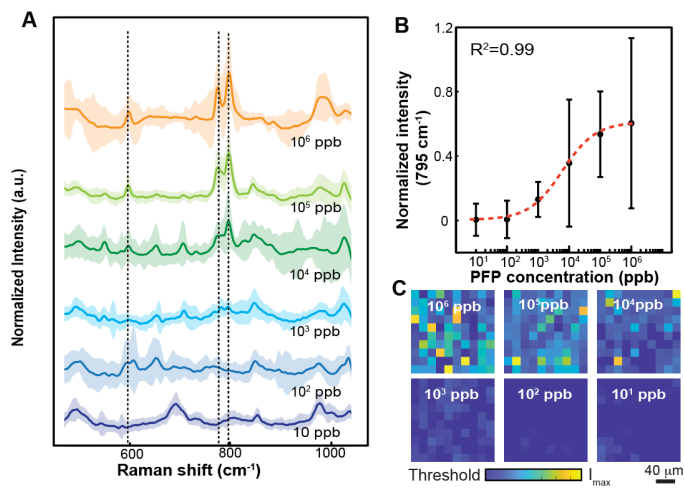

**Figure S4:** Technical replicate for PFP quantification. (A) SERS spectra of volatilized PFP across concentrations ranging from $10^7$ to $10^1$ ppb. Shaded regions indicate standard deviations, and black dotted lines mark the characteristic PFP peaks at 595, 770, and 795 cm$^{-1}$. (B) Quantitative analysis of the 795 cm$^{-1}$ PFP peak intensity across the concentration range (Error bars represent standard deviations; the red curve shows the sigmoidal calibration fit). (C) Spatial SERS intensity maps of the 795 cm$^{-1}$ PFP peak. Threshold is defined as mean plus two times the standard deviation of the control.